\documentclass[journal,final,twocolumn]{IEEEtran}

\usepackage[latin9]{inputenc} \usepackage[T1]{fontenc}

\usepackage{cite,graphicx,url,amsmath}

\usepackage{color,algorithm,alltt}
\usepackage{amsfonts,amssymb}

\newcommand{\nat}{\mathbb{N}}
\newcommand{\real}{\mathbb{R}}
\newcommand{\I}[1]{\mathbf{#1}}
\newcommand{\C}[1]{[#1]}
\newcommand{\Cm}[2]{\C{#1}_{#2}}
\newcommand{\Cn}[1]{\Cm{#1}{n}}
\newcommand{\Gn}[1]{\Cm{#1}{n}^{\Gamma}}

\newcommand{\ub}[1]{\overline{\I{#1}}}
\newcommand{\lb}[1]{\underline{\I{#1}}}
\newcommand{\UB}[1]{\overline{#1}}
\newcommand{\LB}[1]{\underline{#1}}
\newcommand{\IMPLIES}{$\Longrightarrow$}
\newcommand{\impLIES}{\ \Longrightarrow\ }

\newcommand{\Sqrt}{{\rm sqrt}}
\newcommand{\equref}[1]{(\ref{eqn.#1})}
\newcommand{\rat}{\mathbb{Q}}

\newcommand{\atan}{{\rm atan}}
\newcommand{\loatan}[2]{\LB{\atan}({#1},{#2})}
\newcommand{\hiatan}[2]{\UB{\atan}({#1},{#2})}
\newcommand{\lopi}[1]{\LB{\pi}(#1)}
\newcommand{\hipi}[1]{\UB{\pi}(#1)}
\newcommand{\loln}[2]{\LB{\ln}({#1},{#2})}
\newcommand{\hiln}[2]{\UB{\ln}({#1},{#2})}
\newcommand{\loexp}[2]{\LB{\exp}({#1},{#2})}
\newcommand{\hiexp}[2]{\UB{\exp}({#1},{#2})}

\newbox\tempa
\newbox\tempb
\newdimen\tempc
\def\mud#1{\hfil $\displaystyle{\mathstrut #1}$\hfil}
\def\rig#1{\hfil $\displaystyle{#1}$}
\def\irulehelp#1#2#3{\setbox\tempa=\hbox{$\displaystyle{\mathstrut #2}$}%
                        \setbox\tempb=\vbox{\halign{##\cr
        \mud{#1}\cr
        \noalign{\vskip\the\lineskip}
        \noalign{\hrule height 0pt}%
        \rig{\vbox to 0pt{\vss\hbox to 0pt{${\; #3}$\hss}\vss}}\cr
        \noalign{\hrule}%
        \noalign{\vskip\the\lineskip}%
        \mud{\copy\tempa}\cr}}%
                      \tempc=\wd\tempb
                      \advance\tempc by \wd\tempa
                      \divide\tempc by 2 }
\def\irule#1#2#3{{\irulehelp{#1}{#2}{#3}%
                     \hbox to \wd\tempa{\hss \box\tempb \hss}}}

\newcommand\nrule[3]{\irule{#1}{#2}{\ \mbox{\small[#3]}}}

\newcommand{\Cosposn}{5}

\floatname{algorithm}{PVS Listing}

\newcommand{\cesar}[1]{}
\newcommand{\marc}[1]{}
\newcommand{\david}[1]{}

\renewcommand{\cesar}[1]{\textcolor{blue}{\footnotesize [Cesar says: #1]}}
\renewcommand{\marc}[1]{\textcolor{red}{\footnotesize [Marc says: #1]}}
\renewcommand{\david}[1]{\textcolor{magenta}{[David says: #1]}}

\newtheorem{proposition}{Proposition}
\newtheorem{theorem}{Theorem}

\begin{document}

\title{Verified Real Number Calculations: \\
  A Library for Interval 
  Arithmetic}

\author{
  Marc Daumas, David Lester, and~C\'esar Mu\~noz
  \thanks{
    M.  Daumas ({\tt marc.daumas@lirmm.fr}) is with the LIRMM, CNRS,
    UM2 and ELIAUS, UPVD and he is supported in part by the PICS 2533 of the
    CNRS.}
  \thanks{
    D. Lester (\texttt{dlester@cs.man.ac.uk}) is with the University
    of Manchester, Oxford Road, Manchester M13 9PL, UK and he is supported
    in part by the French Région Languedoc-Roussillon.}
  \thanks{
    C. Mu\~noz (\texttt{munoz@nianet.org}) is with the National
    Institute of Aerospace, 100 Exploration Way, Hampton, VA 23666, USA and
    he is supported in part by he National Aeronautics and Space
    Administration under NASA Cooperative Agreement NCC-1-02043 and by
    the University of Perpignan.}
}

\maketitle

\begin{abstract}
  Real number calculations on elementary functions are remarkably
  difficult to handle in mechanical proofs.  In this
  paper, we show how these calculations can be performed within a theorem
  prover or proof assistant in a convenient and highly automated as well as interactive way. 
  First, we formally establish upper and lower
  bounds for elementary functions.  Then, based on these bounds, we
  develop a rational interval arithmetic where real number
  calculations take place in an algebraic setting.  In order to reduce
  the dependency effect of interval arithmetic, we integrate
  two techniques: interval splitting and taylor series
  expansions. This pragmatic approach has been developed, and
  formally verified, in a theorem prover. The formal development
  also includes a set of customizable strategies to automate proofs involving
  explicit calculations over real numbers.
  Our ultimate goal is to
  provide guaranteed proofs of numerical properties with minimal human
  theorem-prover interaction.  

\end{abstract}


\section{Introduction}

Deadly and disastrous failures~\cite{USA.92,Lio.96,GagMcC04} confirm
the shared belief that traditional testing, simulation, and
peer-review are not sufficient to guarantee the correctness of critical
software. {\em Formal Methods} in computer science refers to a set of
mathematical techniques and tools to verify safety properties of  a system design and
its implementation functional requirements.
In
the verification of engineering applications, such as aerospace
systems, it is often necessary to perform explicit calculations with
non-algebraic functions.  Despite all of the developments concerning real
analysis in theorem
provers~\cite{Har98,FlePau2K,Gam99,May01,Got02}, the formal
verification of the correctness of these calculations is not routine.

Take, for example, the formula 
\begin{eqnarray}
\label{eqn.ails}
\frac{3\pi}{180}\ \le\ \frac{g}{v}\tan(\frac{35\pi}{180})\ \ \le \frac{3.1 \pi}{180},
\end{eqnarray}
where $g$ is the gravitational force and $v=250$ knots is the ground
speed of an aircraft. This formula appears in the verification of 
NASA's Airborne Information for Lateral Spacing (AILS) algorithm~\cite{MunCarDowBut03}.
It states that the turn rate of an aircraft flying at ground
speed $v$
with a bank angle of $35^o$ is about $3^o$ per second. A direct proof of
this formula is about a page long and requires the use of several trigonometric
properties.

In many cases the formal checking of numerical calculations
is so cumbersome that the effort seems futile; it 
is then tempting to perform the calculations out of the system,
and introduce the results as axioms.\footnote{As a matter of fact,
the original verification of NASA's AILS algorithm contained several such 
axioms.} However, chances are that the external
calculations will be performed using floating-point 
arithmetic. Without formal checking of the results, we will never be sure
of the correctness of the calculations.

In this paper we present a set of interactive tools to automatically 
prove numerical properties, such as Formula~\equref{ails}, 
within a proof assistant.
The point of departure is a collection of lower and upper bounds for 
rational and non-rational operations. Based on provable properties of
these bounds, we develop a rational interval arithmetic which is amenable
to automation. The series approximations and interval arithmetic presented 
here are well-known. However, to our knowledge, this is the most complete 
formalization in a theorem prover of interval arithmetic that includes 
non-algebraic functions.

Our ultimate goal is to provide guaranteed formal proofs of numerical
properties with minimum human effort. As automated processes are bound
to fail on degenerate cases and waste time and memory on simple ones,
we have designed a set of highly customizable proof strategies. The
default values of the parameters are sufficient in most simple cases.
However, a domain expert can set these parameters to obtain a desired
result, {\em e.g.}, the accuracy of a particular calculation.

This paper merges and extends the results presented in
\cite{DauMelMun05,MunLes05}.  The rest of this document is organized as
follows.  Section~\ref{sec.bounds} defines bounds for elementary
functions.  Section~\ref{sec.arith} presents a rational
interval arithmetic based on these bounds. Section~\ref{sec.alg}
describes a method to prove numerical propositions. The implementation
of this method in a theorem prover is described in Section~\ref{sec.pvs}. Last
section summarizes our work and compares it to related work.

The mathematical development presented in this paper has been written
and fully verified in the Prototype Verification System
(PVS)~\cite{OwrRusSha92}\footnote{PVS is available from
  \url{http://pvs.csl.sri.com}.}. PVS provides a strongly typed
specification language and a theorem prover for higher-order logic. It 
is developed by SRI International.  Our development is freely
available on the Internet.  The results on upper and lower bounds have
been integrated to the NASA Langley PVS
Libraries\footnote{\url{http://shemesh.larc.nasa.gov/fm/ftp/larc/PVS-library/pvslib.html}.}
and the rational interval arithmetic and the PVS strategies for
numerical propositions are available from one of the
authors\footnote{\url{http://research.nianet.org/~munoz/Interval}.}.

For readability, we will use standard mathematical notations along this
paper and PVS notations will be limited to illustrate the use of the library.
In the following, we use the first letters of the alphabet
$a,b,\ldots$ to denote rational numbers, and the last letters of the
alphabet $ \ldots x,y,z$ to denote arbitrary real variables.  We use
{\bf boldface} for interval variables.  Furthermore, if $\I{x}$ is an
interval variable, $\lb{x}$ denotes its lower bound and $\ub{x}$
denotes its upper bound.

\section{Bounds for Elementary Functions}
\label{sec.bounds}
A PVS basic theory of bounds for square root and trigonometric 
functions was originally proposed for the verification of 
NASA's AILS algorithm~\cite{MunCarDowBut03}. 
We have completed it and extended with bounds for 
natural logarithm, exponential, and arctangent. 
The basic idea is to provide for each real function 
$f:\real \mapsto \real$, functions 
$\LB{f}:(\real,\nat)  \mapsto \real$ and
$\UB{f}:(\real,\nat)  \mapsto \real$ closed under $\rat$, 
such that for all $x$, $n$
\begin{eqnarray}
\label{eqn.ublb}
\LB{f}(x,n)\ \le &\ f(x)\ & \le\ \UB{f}(x,n),\\
\label{eqn.lb}
\LB{f}(x,n) &\le& \LB{f}(x,n+1),\\ 
\label{eqn.ub}
\UB{f}(x,n+1) &\le& \UB{f}(x,n),\\
\label{eqn.lim}
\lim_{n\rightarrow \infty} \LB{f}(x,n)\ =&f(x)&=\  
\lim_{n\rightarrow \infty}\UB{f}(x,n).
\end{eqnarray}
Formula~\equref{ublb} states that $\LB{f}$ and $\UB{f}$ are,
respectively, lower and upper bounds of $f$, and
formulas~\equref{lb},~\equref{ub},~and~\equref{lim} state that these
bounds can ultimately be improved, as much as needed, by increasing
the approximation parameter~$n$.

For transcendental functions, we use taylor approximation series.
We
performed a coarse range reduction \cite{AbrSte72} since  the convergence of taylor series  is usually best for small values. More elaborate
range reduction techniques \cite{Mul06} would significantly enhance the speed and
the accuracy of the functions defined in Sections~\ref{sec.bounds} and \ref{sec.arith}.  All the stated propositions in this section
have been formally verified in the verification system PVS.

\subsection{Square root}
For square root, we use a simple approximation by Newton's method. 
For $x \ge 0$,
\begin{eqnarray*}
\UB\Sqrt(x,0)&=&x+1,\\
\UB\Sqrt(x,n+1)&=&\frac{1}{2}(y+\frac{x}{y}),\ \ \
\mbox{where }y = \UB{\Sqrt}(x,n),\\
\LB{\Sqrt}(x,n)&=&\frac{x}{\UB{\Sqrt}(x,n)}.
\end{eqnarray*}

\begin{proposition}
\label{prop.sqrt}
$\forall x \ge 0,n:~0\ \le\ \LB{\Sqrt}(x,n)\ \le\ \sqrt{x}\ <\ 
  \UB{\Sqrt}(x,n).$
\end{proposition}
The first inequality is strict when $x > 0$.

\subsection{Trigonometric functions}
We use the partial approximation by series. 
\begin{eqnarray*}
\LB{\sin}(x,n)&=& \sum_{i=1}^{m} (-1)^{i-1}\frac{x^{2i-1}}{(2i-1)!}\\
\UB{\sin}(x,n)&=& \sum_{i=1}^{m+1} (-1)^{i-1}\frac{x^{2i-1}}{(2i-1)!},\\
\LB{\cos}(x,n)&=& 1 + \sum_{i=1}^{m+1} (-1)^{i}\frac{x^{2i}}{(2i)!},\\
\UB{\cos}(x,n)&=& 1 + \sum_{i=1}^{m} (-1)^{i}\frac{x^{2i}}{(2i)!},\\
\end{eqnarray*}
where $m = 2n$ if $x < 0$; otherwise, $m=2n+1$. 

\begin{proposition}
\label{prop.sin} $\forall x,n:~\LB{\sin}(x,n)\ \le\ \sin(x)\ \le\ \UB{\sin}(x,n)$. 
\end{proposition}

\begin{proposition}
\label{prop.cos} $\forall x,n:~\LB{\cos}(x,n)\ \le\ \cos(x)\ \le\ \UB{\cos}(x,n)$. 
\end{proposition}

\subsection{Arctangent and $\pi$}

We first use the alternating partial approximation by series for $0 \le x \le 1$.
\begin{eqnarray*}
\loatan{x}{n} & = &\sum_{i=1}^{2n+1} x^{2i+1}\frac{(-1)^i}{2i+1},\ \ \
\mbox{if } 0 < x \le 1,\\
\hiatan{x}{n} & = &\sum_{i=1}^{2n} x^{2i+1}\frac{(-1)^i}{2i+1},\ \ \ 
\mbox{if } 0 < x \le 1.
\end{eqnarray*}
We note that for $x=1$ (which we might na\"{\i}vely wish to use to
define $\pi/4$ and hence $\pi$) the series:
$1-\frac{1}{3}+\frac{1}{5}-\frac{1}{7}+\frac{1}{9}-\cdots$
{\em does} converge, but very slowly. Instead, we use the equality
$\frac{\pi}{4} = 4\ \atan(1/5) - \atan(1/239)$,
that has much better convergence properties.
Using this identity we can define bounds on $\pi$:
\begin{eqnarray*}
\lopi{n} &=& 16\ \loatan{1}{n} - 4\ \hiatan{1}{n},\\
\hipi{n} &=& 16\ \hiatan{1}{n} - 4\ \loatan{1}{n}.
\end{eqnarray*}
\begin{proposition}
\label{prop.pi} $\forall n:~ \lopi{n} \ \le\ \pi\ \le\ \hipi{n}$.
\end{proposition}
Now, using properties of arctangent, we extend the range of 
the function to the whole set of real numbers:
\begin{eqnarray*}
\loatan{0}{n} & =& \hiatan{0}{n}~=~0,\\
\loatan{x}{n} & = & \frac{\lopi{n}}{2} - \hiatan{\frac{1}{x}}{n},\ \ \
\mbox{if } 1 < x,\\
\loatan{x}{n} & = & -\hiatan{-x}{n}, \ \ \
\mbox{if } x < 0,\\
\hiatan{x}{n} & = & \frac{\hipi{n}}{2} - \loatan{\frac{1}{x}}{n},\ \ \
\mbox{if } 1 < x,\\
\hiatan{x}{n} & = & -\loatan{-x}{n}, \ \ \
\mbox{if } x < 0.\\
\end{eqnarray*}

\begin{proposition}
\label{prop.atan} $\forall x,n:~\loatan{x}{n}\ \le\ \atan(x)\ 
\le\ \hiatan{x}{n}$.
\end{proposition}
These are strict inequalities except when $x=0$.

The PVS definition of bounds on $\atan$ and $\pi$ are presented in
Listing~\ref{lst.atanpi}. PVS developments
are organized in theories, which are collections of mathematical
and logical objects such as function definitions, variable
declarations, axioms, and lemmas.
The  \texttt{atan\_approx} theory first
imports the definition of the arctangent function. Then, it declares
variables \texttt{n,x,px} of types \texttt{nat} (natural numbers), 
\texttt{real} (real numbers), and \texttt{posreal} (positive real numbers),
respectively. For the scope of the theory, these variables are implicitly 
universally quantified.
Though writing definitions, lemmas, theorems and specially proofs in
PVS requires some training, reading theories is possible to anybody
with a minimal background in logic.

\begin{algorithm}
\caption{Definition of bounds on $\atan$ and $\pi$}
\label{lst.atanpi}\small
\renewcommand{\baselinestretch}{1}\footnotesize %
\begin{alltt}

atan_approx: THEORY
BEGIN

  IMPORTING atan

  n:  VAR nat
  x:  VAR real
  px: VAR posreal

  atan_pos_le1_ub(n,x): real = 
    atan_series_n(x,2*n)

  atan_pos_le1_lb(n,x): real = 
    atan_series_n(x,2*n+1)

  atan_pos_le1_bounds: LEMMA 
    0 < x AND x <= 1 IMPLIES
      atan_pos_le1_lb(n,x) < atan(x) AND 
      atan(x) < atan_pos_le1_ub(n,x)

  pi_lbn(n): posreal = 
    4*(4*atan_pos_le1_lb(n,1/5) - 
       atan_pos_le1_ub(n,1/239))

  pi_ubn(n): posreal = 
    4*(4*atan_pos_le1_ub(n,1/5) - 
       atan_pos_le1_lb(n,1/239))

  pi_bounds: THEOREM 
    pi_lbn(n) < pi AND pi < pi_ubn(n)

  atan_pos_lb(n,px): real = 
    IF px <= 1 THEN 
      atan_pos_le1_lb(n,px) 
    ELSE
      pi_lbn(n)/2 - atan_pos_le1_ub(n,1/px) 
    ENDIF

  atan_pos_ub(n,px): real = 
    IF px <= 1 THEN 
      atan_pos_le1_ub(n,px) 
    ELSE
      pi_ubn(n)/2 - atan_pos_le1_lb(n,1/px) 
    ENDIF

  atan_lb(x,n): real = 
    IF x > 0 THEN atan_pos_lb(n,x)
    ELSIF x = 0 THEN 0
    ELSE -atan_pos_ub(n,-x) ENDIF

  atan_ub(x,n): real = 
    IF x > 0 THEN atan_pos_ub(n,x)
    ELSIF x = 0 THEN 0
    ELSE -atan_pos_lb(n,-x) ENDIF

  atan_bounds: THEOREM
    atan_lb(x,n) <= atan(x) AND 
    atan(x) <= atan_ub(x,n)

END atan_approx
\end{alltt}
\end{algorithm}

\subsection{Exponential}
The series we use for the exponential function is
\begin{eqnarray*}
\exp(x)=\sum_{i=0}^\infty \frac{x^i}{i!}.
\end{eqnarray*}
We could directly find bounds for negative $x$ from this series as,
in this case, the series is alternating.
However, we will subsequently find that it is convenient to show that our
bounds for the exponential function are strictly positive, 
and this is not
true for all $x \le 0$. Yet, this property  {\em holds}  for $-1 \le x \le 0$. 

We define
\begin{eqnarray*}
\loexp{x}{n} & = &\sum_{i=0}^{2(n+1)+1} \frac{x^i}{i!},\ \ \
\mbox{if } -1 \le x < 0,\\
\hiexp{x}{n} & = &\sum_{i=0}^{2(n+1)} \frac{x^i}{i!},\ \ \
\mbox{if } -1 \le x < 0.
\end{eqnarray*}

Using properties of the exponential function, we 
obtain bounds for the whole set of real numbers:
\begin{eqnarray*}
\loexp{0}{n} &=& \hiexp{0}{n}~=~1,\\
\loexp{x}{n} &=&{\loexp{\frac{x}{-\lfloor x\rfloor}}{n}}^{-\lfloor x \rfloor},\ \ \
\mbox{if } x < -1\\
\loexp{x}{n} &=& \frac{1}{\hiexp{-x}{n}},\ \ \
\mbox{if } x > 0,\\
\hiexp{x}{n} &=&{\hiexp{\frac{x}{-\lfloor x \rfloor}}{n}}^{-\lfloor x \rfloor},\ \ \
\mbox{if } x < -1\\
\hiexp{x}{n} &=& \frac{1}{\loexp{-x}{n}},\ \ \
\mbox{if } x > 0.\\
\end{eqnarray*}
Notice that unless we can ensure that all of the bounding functions
are strictly positive we will run into type-checking problems using the 
bound definitions for $x > 0$, {\em e.g.}, $1/\hiexp{-x}{n}$ is only defined provided
$\hiexp{-x}{n} \not=0$.

\begin{proposition}
\label{prop.exp}$\forall x,n:~0\ <\ \loexp{x}{n}\ \le\ \exp(x)\ \le\ \hiexp{x}{n}$.
\end{proposition}
These are strict inequalities except when $x=0$. 

\subsection{Natural Logarithm}
For $0 < x \le 1$, we use the alternating series for natural 
logarithm:
\begin{eqnarray*}
\ln(x+1) &=&\sum_{i=1}^{\infty} (-1)^{i+1} \frac{x^i}{i}.
\end{eqnarray*}
Therefore, we define
\begin{eqnarray*}
\loln{x}{n} & = &\sum_{i=1}^{2n} (-1)^{i+1}\frac{(x-1)^i}{i},\ \ \ 
\mbox{if }1 < x \le 2,\\
\hiln{x}{n} & = &\sum_{i=1}^{2n+1} (-1)^{i+1}\frac{(x-1)^i}{i},\ \ \
\mbox{if }1 < x \le 2.
\end{eqnarray*}
Using properties of the natural logarithm function, we obtain
\begin{eqnarray*}
\loln{1}{n} & = & \hiln{1}{n}~=~0\\
\loln{x}{n} & = & -\loln{\frac{1}{x}}{n},\ \ \
\mbox{if } 0 < x < 1,\\
\hiln{x}{n} & = & -\hiln{\frac{1}{x}}{n},\ \ \
\mbox{if } 0 < x < 1.\\
\end{eqnarray*}
Finally, we extend the range to the whole set of positive reals. 
If $x > 2$, we find a natural number $m$ and real number $y$
such that $x = 2^my$ and $1 < y \le 2$, by using the following 
recursive algorithm similar in spirit to Euclidean division:
\begin{alltt}\small
lnnat(x:posreal,k:posnat): [nat,posreal] =
  if x < k then (0,x)
  else 
    let (m,y) = lnnat(x/k,k) in 
    (m+1,y)
  endif
\end{alltt}
We next prove the following property:
\begin{proposition}
\label{prop.lnnat}
$\forall x\ge1, k>1:~k^m \le x < k^{m+1}, y < k, x = k^m y$,
where $(m,y) = \texttt{lnnat}(x,k)$.
\end{proposition}

 If $(m,y) = \texttt{lnnat}(2,x)$, we observe that 
\[\ln(x)=\ln(2^my)=m\ln(2) + \ln(y).\]
Hence,
\begin{eqnarray*}
\loln{x}{n} & = &m\ \loln{2}{n} + \loln{y}{n},\ \ \
\mbox{if } x > 2,\\
\hiln{x}{n} & = &m\ \hiln{2}{n} + \hiln{y}{n},\ \ \
\mbox{if } x > 2.
\end{eqnarray*}

\begin{proposition}
\label{prop.log}
$\forall x>0,n:~\loln{x}{n}\ \le\ \ln(x)\ \le\ \hiln{x}{n}$.
\end{proposition}
These are strict inequalities except when $x=1$.

\section{Rational Interval Arithmetic}
\label{sec.arith}
Interval arithmetic has been used for decades as a standard tool for
numerical analysis on engineering applications
\cite{Neu90,JauKieDidWal01}. 
In interval arithmetic, operations are evaluated on range of numbers 
rather than on real numbers. A {\em (closed) interval} $\C{a,b}$ is the set of real 
numbers between~$a$ and~$b$, {\em i.e.},
\begin{eqnarray*}
\C{a,b} &=& \{ x\ |\ a \le x \le b \}.
\end{eqnarray*} 
The bounds $a$ and $b$ are called 
the {\em lower bound} and {\em upper bound} of $\C{a,b}$, respectively. 
Note that if $a > b$, the interval is the empty set. 
The notation $\C{a}$ 
abbreviates the point-wise interval $\C{a,a}$. 

Interval computations can be performed on the endpoints or on the
center and the radius.  For this work, we decided to work on rational
endpoints.  Trigonometric and transcendental functions for interval
arithmetic are defined using the bounds presented in
Section~\ref{sec.bounds}. %

Listing~{\ref{lst.defs}} shows a few definitions from the PVS theory
\texttt{Interval}. Dots are used to simplify the presentation and hide
some technical parts.
The theory defines the type \texttt{Interval} as a record with fields
\texttt{ub} and \texttt{lb} of type \texttt{rat} (rational numbers), 
variables \texttt{x,y} of type
\texttt{real}, variable \texttt{n} of type \texttt{nat}, and variables
\texttt{X,Y} of type \texttt{Interval}.

\begin{algorithm}
  \caption{Definition of interval arithmetic}
  \label{lst.defs}
   \begin{alltt}\small

Interval : THEORY
BEGIN

  Interval : TYPE = [#
               lb : rat,
               ub : rat
             #]

  x,y : VAR real
  n   : VAR nat
  X,Y : VAR Interval

  +(X,Y): Interval = [|lb(X)+lb(Y),
                       ub(X)+ub(Y)|]
  -(X,Y): Interval = [|lb(X)-ub(Y),     
                       ub(X)-lb(Y)|]
  -(X)  : Interval = [|-ub(X), 
                       -lb(X))|]
  *(X,Y): Interval = \ldots
  /(X,Y): Interval = X * [|1/ub(Y), 
                           1/lb(Y)|]
  Abs(X): Interval = \ldots
  Sq(X) : Interval = \ldots
  ^(X,n): Interval = \ldots

  U(X,Y) : Interval = [|min(lb(X),lb(Y)),
                        max(ub(X),ub(Y))|]
  \ldots 

END Interval   
\end{alltt}
\end{algorithm}

If \texttt{X} is
a PVS interval, \texttt{lb(X)} is the lower bound and \texttt{ub(X)}
is the upper bound of \texttt{X}.  In PVS, we define the syntactic
sugar \texttt{[|x,y|]} to represent the interval $[x,y]$. 
Interval union $\I{x} \cup \I{y}$, written in PVS \texttt{X
U Y}, is defined as the smallest rational interval that contains both
$\I{x}$ and $\I{y}$.

The four basic interval operations are defined as follows~\cite{Kea96}:
\begin{eqnarray*}
\I{x} + \I{y} &=& \C{\lb{x}+\lb{y},\ub{x}+\ub{y}},\\
\I{x} - \I{y} &=& \C{\lb{x}-\ub{y},\ub{x}-\lb{y}},\\
\I{x} \times \I{y} &=& 
\C{\min\{\lb{x}\lb{y},\lb{x}\ub{y},\ub{x}\lb{y},\ub{x}\ub{y}\},
 \max\{\lb{x}\lb{y},\lb{x}\ub{y},\ub{x}\lb{y},\ub{x}\ub{y}\}},\\
\I{x} / \I{y} &=& \I{x} \times \C{\frac{1}{\ub{y}},\frac{1}{\lb{y}}},\ \ \ 
\mbox{if } \lb{y}\ub{y} > 0.
\end{eqnarray*}
We also define the unary negation, absolute value, and power operators 
for intervals:
\begin{eqnarray*}
-{\I{x}} &=& \C{-\ub{x},-\lb{x}},\\
|\I{x}| &=& \C{\min\{|\lb{x}|,|\ub{x}|\},\max\{|\lb{x}|,|\ub{x}|\}},\ \ \
\mbox{if } \lb{x}\ub{x} \ge 0.\\
|\I{x}| &=& \C{0,\max\{|\lb{x}|,|\ub{x}|\}},\ \ \
\mbox{if } \lb{x}\ub{x} < 0.\\
\I{x}^n &=& \left\{
\begin{array}{l@{\ \ }l}
\C{1}&\mbox{if }n=0, \\
\C{\lb{x}^n,\ub{x}^n}&\mbox{if }\lb{x} \ge 0\ \mbox{ or } \mbox{odd?}(n),\\
\C{\ub{x}^n,\lb{x}^n}&\mbox{if }\ub{x} \le 0\ \mbox{ and } \mbox{even?}(n),\\
\C{0,\max\{\lb{x}^n,\ub{x}^n\}}&\mbox{otherwise}.
\end{array}\right.
\end{eqnarray*}

Interval operations are defined such that they include the result of
their corresponding real operations. This property is called the {\em
inclusion property}. 
\begin{proposition}[Inclusion Property for Basic Operators]
\label{prop.basic}
If $x \in \I{x}$ and $y \in \I{y}$ then 
$x \otimes y \in \I{x} \otimes \I{y}$, where 
$\otimes \in \{+,-,\times,/\}$. Moreover, $-x \in -\I{x}$, 
$|x| \in |\I{x}|$, and 
$x^n \in \I{x}^n$, for $n \ge 0$.
It is assumed that $\I{y}$ does not contain $0$ in the case of interval
division. 
\end{proposition}
Listing~{\ref{lst.props}} specifies this property in PVS.
The
proposition $x \in \I{x}$ %
is written \texttt{x~\#\#~X}. %

\begin{algorithm}
  \caption{Basic inclusion properties}
  \label{lst.props}
   \begin{alltt}\small

  Add_inclusion : LEMMA
    x ## X AND y ## Y \IMPLIES x+y ## X+Y

  Sub_inclusion : LEMMA
    x ## X AND y ## Y \IMPLIES x-y ## X-Y

  Neg_inclusion : LEMMA
    x ## X \IMPLIES -x ## -X

  Mult_inclusion : LEMMA
    x ## X AND y ## Y \IMPLIES x*y ## X*Y

  Div_inclusion : LEMMA
    NOT 0 ## Y AND
    x ## X AND y ## Y \IMPLIES x/y ## X/Y

  Abs_inclusion : LEMMA
    x ## X \IMPLIES abs(x) ## abs(X)

  Sq_inclusion : LEMMA
    x ## X \IMPLIES sq(x) ## sq(X)

  Pow_inclusion : LEMMA
    x ## X \IMPLIES x^n ## X^n
\end{alltt}
\end{algorithm}

The inclusion property is fundamental to interval arithmetic. It
guarantees that evaluations of an expression using interval
arithmetic bound its  exact real value. Any operation in interval 
arithmetic must satisfy the inclusion property with respect to
its corresponding real operation. 

\subsection{Interval comparisons}
\label{sec.bow}
There are several possible ways to compare intervals~\cite{Yak92}. 
In this work, we use interval-rational comparisons and interval inclusions.
\begin{eqnarray*}
\I{x} &<& a\ \ \ \mbox{if } \ub{x} < a, \mbox{ similarly for $\le$},\\
\I{x} &>& a\ \ \ \mbox{if } \lb{x} > a, \mbox{ similarly for $\ge$},\\
\I{x} &\subseteq& \I{y}\ \ \
\mbox{if } \lb{y} \le \lb{x} \mbox{ and } \ub{x} \le \ub{y}.
\end{eqnarray*}

\begin{proposition}
Assume that $x \in \I{x}$,
\begin{enumerate}
\label{prop.bow}
\item if $\I{x} \bowtie a$ then $x \bowtie a$, for
$\bowtie\ \in \{<,\le,>,\ge\}$, and
\item if $\I{x} \subseteq \I{y}$ then $x \in \I{y}$.
\end{enumerate}
\end{proposition}

We use $\not\bowtie$ to denote $\ge$, $>$, $\le$, or $<$, when 
$\bowtie$ is, respectively, $<$, $\le$, $>$, or $\ge$.
\begin{proposition}
\label{prop.notbow}
If $\I{x} \bowtie a$ and $\I{x} \not\bowtie a$, then $\I{x}$ is empty.
\end{proposition}
Notice that $\neg (\I{x} \bowtie a)$ does not imply $\I{x} \not\bowtie a$.
For instance, $\C{-1,1}$ is neither greater nor less than $0$.

\subsection{Square root, arctangent, exponential, and natural logarithm}
Interval functions for square root, arctangent, $\pi$, 
exponential, and natural logarithm are defined for an approximation
parameter $n \ge 0$:
\begin{eqnarray*}
\Cn{\sqrt{\I{x}}}&=&\C{\LB{\Sqrt}(\lb{x},n),\UB{\Sqrt}(\ub{x},n)}, \ \ \
\mbox{if } \I{x} \ge 0,\\
\Cn{\atan({\I{x}})}&=&\C{\LB{\atan}(\lb{x},n),\UB{\atan}(\ub{x},n)},\\
\Cn{\pi}&=&\C{\LB{\pi}(n),\UB{\pi}(n)},\\
\Cn{\exp({\I{x}})}&=&\C{\LB{\exp}(\lb{x},n),\UB{\exp}(\ub{x},n)},\\
\Cn{\ln({\I{x}})}&=&\C{\LB{\ln}(\lb{x},n),\UB{\ln}(\ub{x},n)},\ \ \
\mbox{if } \I{x} > 0.
\end{eqnarray*}

As consequence of Propositions~\ref{prop.sqrt},~\ref{prop.atan},
~\ref{prop.exp},~and~\ref{prop.log} in Section~{\ref{sec.bounds}}, and
the fact that these functions are increasing,  
the above functions satisfy the following inclusion property.
\begin{proposition}
\label{prop.trans}
For all $n$, if $x \in \I{x}$ then $f(x) \in \Cn{f(\I{x})}$, where 
$f \in \{\sqrt{\ },\atan,\exp,\ln\}$.  Moreover, $\pi \in \Cn{\pi}$.
It is assumed that $\I{x}$ is non-negative in the case of square root, and
$\I{x}$ is positive in the case of natural logarithm.
\end{proposition}

\subsection{Trigonometric functions}
\label{sec.trig}
Parametric functions for interval trigonometric functions
are defined by cases analysis on quadrants where the functions are increasing
or decreasing. The mathematical definitions are presented in 
Figure~\ref{fig/trig}.

\begin{figure*}
  \begin{eqnarray}
    \label{eqn.sin}
    \Cn{\sin(\I{x})} &=& \left\{
      \begin{array}{l@{\ \ }r l }
        \C{\LB{\sin}(\lb{x},n),\UB{\sin}(\ub{x},n)}            & \mbox{if      } & \I{x} \subseteq \C{-\frac{\lopi{n}}{2},\frac{\lopi{n}}{2}},\\
        \C{\LB{\sin}(\ub{x},n),\UB{\sin}(\lb{x},n)}            & \mbox{else if } & \I{x} \subseteq \C{\frac{\hipi{n}}{2} ,\lopi{n}},\\
        \C{\min\{\LB{\sin}(\lb{x},n),\LB{\sin}(\ub{x},n)\},1}  & \mbox{else if } & \I{x} \subseteq \C{0                  ,\lopi{n}},\\
        -\Cn{\sin(-\I{x})}                                     & \mbox{else if } & \I{x} \subseteq \C{-\lopi{n},0},\\
        \C{-1,1}                                               &\mbox{otherwise},  
      \end{array}
    \right.\\
    \label{eqn.cos}
    \Cn{\cos(\I{x})} &=& \left\{
      \begin{array}{l@{\ \ }r l}
        \C{\LB{\cos}(\ub{x},n),\UB{\cos}(\lb{x},n)}          & \mbox{if      } & \I{x} \subseteq \C{0,\lopi{n}},\\
        \Cn{\cos(-\I{x})}                                    & \mbox{else if } & \I{x} \subseteq \C{-\lopi{n},0},\\
        \C{\min\{\LB{\cos}(\lb{x},n),\LB{\cos}(\ub{x},n)\},1}& \mbox{else if } & \I{x} \subseteq \C{-\frac{\lopi{n}}{2},\frac{\lopi{n}}{2}},\\
        \C{-1,1}                                             & \mbox{otherwise},
      \end{array}
    \right. \\
    \label{eqn.tan}
    \Cn{\tan(\I{x})} &=& 
    \C{\frac{\LB{\sin}}{\UB{\cos}}(\lb{x},n+\Cosposn),\frac{\UB{\sin}}{\LB{\cos}}(\ub{x},n+\Cosposn)},\ \ \
    \mbox{if }\I{x} \subseteq \C{-\frac{\lopi{n+\Cosposn}}{2},\frac{\lopi{n+\Cosposn}}{2}}.
  \end{eqnarray}
  \caption{Interval trigonometric functions}
  \label{fig/trig}
\end{figure*}

Note that $\sin$ and $\cos$ are defined for the whole real line. However,
for angles $\alpha$ such that $|\alpha| > \LB{\pi}$ both functions will
return the interval $\C{-1,1}$, a valid bound but not a very good one.
Furthermore, 
the expression $n+\Cosposn$ in 
Formula~\equref{tan} is necessary to guarantee that 
lower and upper bounds of cosine are strictly positive in the interval
$\C{-\frac{\lopi{n+\Cosposn}}{2},\frac{\lopi{n+\Cosposn}}{2}}$, and thus,
the interval tangent function is always defined in that interval.

The interval trigonometric functions
satisfy the inclusion property.
\begin{proposition}
\label{prop.trig}
If $x \in \I{x}$ then $f(x) \in \Cn{f({\I{x}})}$, where
$f \in \{\sin,\cos\}$. Moreover, 
if $\I{x} \subseteq \C{-\frac{\lopi{n+\Cosposn}}{2},\frac{\lopi{n+\Cosposn}}{2}}$,  
$\tan(x) \in \Cn{\tan(\I{x})}$.
\end{proposition}

\bigskip
The next section proposes a method to prove numerical propositions
based on the interval arithmetic described here.

\section{Mechanical Proofs of Numerical Propositions}
\label{sec.alg}

{\em Arithmetic expressions} are defined by the following grammar, where 
$\cal V$ is an denumerable set of real variables:
\[\begin{array}{lcl}
e &\ ::=\  & a \ \mid\ x \ \mid\ e + e\ \mid\ e - e\ \mid\ -e\ \mid\ e \times e\ \mid\ \\
  &        & e / e\ \mid\ |e| \ \mid\ e^i\ \mid\ \sqrt{e}\ \mid \pi\ \mid\ \sin(e)\ \mid\ \\
  &        & \cos(e)\ \mid\ \tan(e)\ \mid\ \exp(e) \mid\ \ln(e)\mid\ \atan(e)\\
a & \in    & \rat \\
i & \in    & \nat \\
x & \in    & {\cal V}
\end{array}\]
Numerical propositions $P$ have either the form $e_1 \bowtie e_2$, where 
$\bowtie \ \in \{<,\le,>,\ge\}$, or the form $e \in \I{a}$, where $\I{a}$
is a constant interval (an interval with constant rational endpoints). 
As usual, parentheses are used to group real and interval 
expressions as needed.

A {\em context}\ $\Gamma$ is a set of hypotheses of the form $x \in \I{x}$.
A {\em ground context} is a context where all the intervals are constant.
In the following, we use logical judgments in the sequent calculus style,
{\em e.g.}, $\Gamma\ \vdash\ P$, where all free variables occurring in $P$ are 
in $\Gamma$. The intended semantics of a judgment 
$\Gamma\ \vdash\ P$ is that the numerical proposition $P$ is true under
the hypotheses~$\Gamma$. 

Given a context $\Gamma$, an approximation parameter $n$, and an 
expression $e$, such that the free variables
of $e$ are in $\Gamma$, we define the interval expression $\Gn{e}$ by 
recursion on $e$.
\begin{eqnarray*}
\Gn{a}&\ =\ & [a],\\
\Gn{x}&=& \I{x},\ \mbox{where } (x \in \I{x}) \in \Gamma,\\
\Gn{e_1 \otimes e_2} &=& \Gn{e_1} \otimes \Gn{e_2},\ \ \ 
\mbox{where } \otimes \in \{+,-,\times,/\},\\
\Gn{e^i} &=&(\Gn{e})^i,\\
\Gn{-e}  &=&-\Gn{e},\\
\Gn{|e|} &=&|\Gn{e}|,\\
\Gn{\pi} &=&\Cn{\pi},\\
\Gn{f(x)}&=&\Cn{f(\Gn{x})}, \\
         &  &\mbox{where } f \in \{\sin,\cos,\tan,\exp,\ln,\atan\}.
\end{eqnarray*}

\begin{theorem}[Inclusion]
\label{theo.incl}
Let $\Gamma$ be a context,  $n$ an approximation parameter, and
$e$ a well-defined arithmetic expression in 
$\Gamma$, {\em i.e.}, 
side conditions for division, square
root, logarithm, and tangent are satisfied,
\begin{eqnarray}
\Gamma\ \vdash\ e \in \Gn{e}.
\end{eqnarray}
\end{theorem}
\begin{proof}
By structural induction on $e$ and propositions~{\ref{prop.pi}},~%
{\ref{prop.basic}},~\ref{prop.trans},~and~\ref{prop.trig}.
\end{proof}

\subsection{A general method for numerical propositions}
\label{sec.method}
We propose a general method to prove numerical propositions.
First, consider a judgment of the form
\begin{eqnarray*}
\Gamma\ \vdash\ e_1 \bowtie e_2,
\end{eqnarray*}
where $\Gamma$ is a ground context.
\begin{enumerate}
\item Select an approximation parameter $n$.
\item Define $e = e_1 - e_2$.
\item \label{go.eval}
      Evaluate $\Gn{e} \bowtie 0$. If it evaluates to true, 
      the following judgment holds
      \begin{eqnarray*}
      \Gamma &\vdash&\Gn{e} \bowtie 0.
      \end{eqnarray*} 
      In that case go to step~\ref{go.next}.
\item Evaluate
      $\Gn{e} \not\bowtie 0$. If this evaluates to true then fail. 
      By Proposition~\ref{prop.notbow}, the judgment 
      $\Gamma \ \vdash\ \Gn{e} \bowtie 0$ cannot hold. If 
      $\Gn{e} \not\bowtie 0$ evaluates to false, increase
      the approximation parameter and return to step~\ref{go.eval}. 
\item \label{go.next}
      By Theorem~\ref{theo.incl},
      \begin{eqnarray*}
      \Gamma &\vdash& e \in \Gn{e}.
      \end{eqnarray*} 
\item Proposition~\ref{prop.bow} yields
      \begin{eqnarray*}
      \Gamma &\vdash& e \bowtie 0.
      \end{eqnarray*} 
\item By definition, 
      \begin{eqnarray*}
      \Gamma &\vdash& e_1 - e_2 \bowtie 0.
      \end{eqnarray*} 
\item Therefore, 
      \begin{eqnarray*}
      \Gamma &\vdash& e_1 \bowtie e_2.
      \end{eqnarray*} 
\end{enumerate}

The method above can be easily adapted to judgments of the form 
$\Gamma\ \vdash\ e \bowtie \I{a}$. In this case, the
interval expression $\Gn{e} \subseteq \I{a}$ is evaluated.
If the expression evaluates to true, then the original judgment holds
by Theorem~{\ref{theo.incl}} and Proposition~{\ref{prop.bow}}. Otherwise,
the method should fail.

The general method is {\em sound}, {\em i.e.}, all the steps
can be effectively computed and each one is formally justified. 
In particular, 
the propositions $\Gn{e} \bowtie 0$, $\Gn{e} \not \bowtie 0$,
and  $\Gn{e} \subseteq \I{a}$
can be mechanically computed as they only
involve rational arithmetic and constant numerical values. 
The method is not {\em complete} as
it does not necessarily terminate. Even if $e$ only involves
the four basic operations and no variables, 
it may be that both $\Gn{e}\bowtie 0$ and 
$\Gn{e}\not\bowtie 0$ evaluate to false.

The absence of a completeness result is a fundamental limitation on
any general computable arithmetic. At a practical level, the problem
arises because all we have available are a sequence of approximations
to the real numbers $x$ and $y$; provided $x$ and $y$ differ, with
luck we will eventually have a pair of approximations whose intervals
do not overlap, and hence we can return a result for $x\bowtie
y$. However, if $x$ and $y$ are the same real number (note we might
not necessarily get the same sequence of approximations for both $x$
and $y$), we can never be sure whether further evaluation might result
in us being able to distinguish the numbers. 

\subsection{Dependency effect}
The {\em dependency effect} is a well-known behavior of interval arithmetic
due to the fact that interval identity is lost in interval evaluations.
This may have surprising results, for
instance $\I{x} - \I{x}$ is $\C{0}$ only if $\I{x}$ is point-wise. Moreover,
as we have seen in Section~\ref{sec.bow}, both $\I{x} \ge a$ and
$\I{x} < a$ may be false.
Additionally, interval
arithmetic is subdistributive, {\em i.e.}, $\I{x}\times (\I{y}+\I{z})
\subseteq \I{x}\times \I{y} + \I{x} \times \I{z}$.  In the general
case the inclusion is strict and some dependency effects appear as soon as a
variable is used more than once in an expression.

For the method presented in Section~~\ref{sec.method}, it means that the arrangement of the expression $e$ matters.
For instance, assume that we want to prove $x \in \C{0,1}\ \vdash\ 2\times x
\ \ge\ x$. This is pretty
obvious in arithmetic as $x$ is a non-negative real. Using our method,
we first consider the arithmetic expression $e=2\times x - x$ and then 
construct
the interval expression $\Gn{e}=2\times \I{x} - \I{x}$, where
$\I{x}=\C{0,1}$. For any approximation parameter $n$, $\Gn{e}$ evaluates to
$\C{-1,2}$ which is neither greater nor less than $0$. Therefore, the
method will not terminate.  
On the other hand, if instead of the arithmetic expression $2\times x
- x$, we consider the equivalent arithmetic expression $x$, we have
$\Gn{x} = \C{0,1}$ and $\C{0,1} \ge 0$ evaluates to true.

A second observation is that because of the dependency effect the width
of intervals also matters. Consider again the expression $e=2 \times x - x$.
We have seen that the interval evaluation of $\Gn{e}$, for $x \in \C{0,1}$,
results in $\C{-1,2}$,
which is not sufficient to prove that $\Gn{e} \ge 0$. On the other hand, the 
expression $\Gn{e}$ 
evaluates to $\C{-1/2,1}$ when $x \in \C{0,1/2}$ and it evaluates to
$\C{0,3/2}$ when $x \in \C{1/2,1}$. Therefore, we can prove that,
for $x \in \C{0,1}$,
$\Gn{e} \subseteq \C{-1/2,1} \cup \C{0,3/2}$, {\em i.e.}, 
$\Gn{e} \subseteq \C{-1/2,3/2}$, 
which is a better approximation than $\C{-1,2}$. If we continue dividing
the interval $\C{0,1}$ and computing the union of the resulting intervals, 
we can eventually prove that $\Gn{e} + \epsilon \ge 0$ 
for an arbitrary small $\epsilon > 0$.

These observations lead to two enhancements of the general method. First,
we could divide each interval in $\Gamma$ before applying 
the general technique. Second, we may want to replace the original 
expression by an equivalent one that is less prone to the dependency effect.

\subsection{Interval splitting}
\label{sec.splitting}
In interval arithmetic, the dependency effect of the union of the parts 
is less than the dependency effect of the whole. Indeed, 
the simplest way to reduce the 
dependency effect is to divide the interval variables into several
tiles (subintervals) and to evaluate the original expression on these 
tiles separately. This technique is called {\em interval splitting} or {\em sub-paving} and
is expressed by the following deduction rule.

\begin{proposition}
\label{prop.splitting}
Let $\Gamma$ be a context, $e$ an expression whose free variables are $x$ and 
those in $\Gamma$, $\I{e}$ an interval expression, and $\I{x},\I{x}_1,\ldots,\I{x}_n$ intervals 
such that $\I{x} = \bigcup_{1 \le i \le n} \I{x}_i$,
\[\nrule{\forall~1 \le i \le n:~x \in \I{x}_i, \Gamma \ \vdash\ e \in \I{e}}
        {x \in \I{x}, \Gamma \ \vdash\ e \in \I{e}}{Splitting}
\]
\end{proposition}

The Splitting rule can be iterated to obtain a splitting for multiple
variables. Note that the number of tiles generated by interval
splitting is exponential in the number of variables. Indeed, if $k_1$
is the number of tiles of the first variable alone, $k_2$ is the number of
tiles of the second variables alone, and so forth, the total number of tiles
to be considered for $m$ variables is $\prod_{1 \le j \le m} k_j$.

The integration of the Splitting rule into 
the general method is straightforward.
First, a splitting is computed for a given set of variables in 
$\Gamma$. Then, the general method is applied to all cases. 
If the general method
is successful in all of them, by Proposition~\ref{prop.splitting},
the original judgment holds. Otherwise, the method fails and a new splitting
may be considered.

\subsection{Taylor Series Expansions}
\label{sec.taylor}
Replacing $2\times x - x$ by $x$ can be done automatically. In fact, as we
will see in Section~\ref{sec.pvs}, these kinds of 
simplifications are performed by our PVS implementation of the general
method. However, these simplifications may not be sufficient even for
simple expressions such as $x \times (1 - x)$, where 
$x \in \C{0.1}$. The subdistributivity property of interval arithmetic states
that the interval evaluation of $x \times (1 - x)$  is better than that
of the equivalent expression $x - x^2$. Unfortunately, that evaluation is
not good enough to prove that $x \times (1 - x) \in \C{0,1/4}$. In this case,
as a domain expert knows, 
the optimal answer is obtained with the equivalent expression 
$1/4 - (1/2 - x)^2$. The solution is a lot less intuitive when non-algebraic
functions are involved.

Taylor's theorem states that a $n$-differentiable function can be 
approximated near a given point by a polynomial of degree $n$ 
whose coefficients depend on 
the derivatives of the function at that point.
In interval 
arithmetic, taylor's theorem can be expressed by the following deduction rule.
\begin{proposition} 
\label{prop.taylor}
Let $\I{x},\I{x}_0,\ldots,\I{x}_n$ be strictly proper intervals, 
$f$ a $n$-differentiable function on a variable $x \in \I{x}$, and 
$c \in \I{x}$ a constant,
\[\nrule
{\begin{array}{c}
\forall~0 \le i < n:\ \vdash\ f^{(i)}(c) \in \I{x}_i\\
x \in \I{x}\ \vdash\ f^{(n)}(x) \in \I{x}_n
\end{array}}
{x \in \I{x}\ \vdash\ f(x) \in \Sigma_{i=0}^{n}(\I{x}_i \times (\I{x}-c)^i)/i!}
{Taylor}
\]
\end{proposition}

\smallskip

The expression of Taylor's rule shows that interval $\I{x}$ appears
only once in each term of order $i$ for $i$ between $1$ and $n - 1$
preventing any dependency effect due to $\I{x}$ in a term alone.  The
term of order $n$ suffers some dependency effect as $\I{x}$ also
appears in the definition of $\I{x}_n$. In most cases, $n = 2$ is used
to cancel first order dependency effects as presented
Listing~\ref{lst.fairatan}. But in cases where the first derivatives
nearly vanish or  where the evaluation of the last derivative
introduces significant dependency effects, we compute more  terms to
reach some better bounds.

Using Taylor's rule require  more
work  than the Splitting rule. In particular, we need 
to provide intervals $\I{x}_0,\ldots,\I{x}_n$ and constant $c$ that satisfy
the hypotheses of the rule. For $c$ we choose the middle point of $\I{x}$ unless the user proposes another point. It follow immediately that  $c \in \I{x}$. For $0 \le i < n$, we choose 
$\I{x}_i = \Cn{f^{(i)}(c)}$ and, by Theorem~\ref{theo.incl}, we have
$f^{(i)}(c) \in \I{x}_i$. Finally, we choose 
$\I{x}_n = \Gn{f^{(n)}(x)}$, where $\Gamma$ is the context 
$x\in\I{x}$. 
By Theorem~\ref{theo.incl}, we have $\Gamma \ \vdash\ f^{(n)}(x) \in \I{x}_n$.

In order to prove the judgment 
$x \in \I{x} \ \vdash\ f(x) \in \I{a}$, we consider the interval expression
$\Sigma_{i=0}^{n}(\I{x}_i \times (\I{x}-c)^i)/i! \subseteq \I{a}$ for a given
$n$. If it evaluates to true, then the original judgment holds 
by Taylor's rule and Proposition~\ref{prop.bow}. If the evaluation
returns false, the method fails and a higher expansion degree $n$ 
may be considered.

For better results, the evaluation of 
$\Sigma_{i=0}^{n}(\I{x}_i \times (\I{x}-c)^i)/i! \subseteq \I{a}$ 
can be performed using the splitting technique.
Contrary to the approach described in~\cite{Saw02}, we do not have to
generate a new taylor approximation for each tile. By using an
interval-based taylor expansion, the same expression can be reused for
all the tiles. One single global taylor expansion has
to be validated, and the proofs for all the tiles simply consist in
an interval evaluation of this expansion. We do not suffer from the taylor coefficients being
irrational numbers, they are simply given by interval expressions
involving rational functions. Relying on rational interval arithmetic leads
to conceptually simpler proofs.

\bigskip
Section~\ref{sec.pvs} describes how the general method and its extensions
are implemented in the PVS theorem prover and illustrates the practical
use of the library with a few examples.

\section{Verified Real Number Calculations in PVS}
\label{sec.pvs}
The interval arithmetic presented in this paper has been
developed as a PVS library called {\tt Interval}. This library
contains the specification of interval arithmetic described here and the
formal proofs of its properties. We believe that a domain expert can use this
library with a basic knowledge of theorem provers. Minimal PVS expertise
is required as most of the technical burden of proving numerical
properties is already implemented as proof strategies.

\subsection{Strategies}
The \texttt{numerical} strategy is the basic strategy that implements
the general method and its extensions described in
Section~\ref{sec.alg}.  For instance, Formula~\ref{eqn.ails} can be
specified in PVS as follows (comments in PVS start with the symbol
\texttt{\%} and extend to the end of the line):
\begin{alltt}\small
  g : posreal = 9.8        %[m/s^2]
  v : posreal = 250*0.514  %[m/s]

  tr35: LEMMA 
    (g*tan(35*pi/180)/v) * 180/pi 
      ## [| 3, 3.1 |]

%|- tr35: PROOF (numerical) QED
\end{alltt}
We emphasize that, in PVS, \texttt{tan} and \texttt{pi} are the real 
mathematical function $\tan$ and constant $\pi$, respectively.
Lemma \texttt{tr35} is automatically discharged by the \texttt{numerical} strategy, which can be entered interactively or in batch
mode, as in this case, via the ProofLite library developed by one of the
authors~\cite{Mun07NIA}.

Another example is the proof of the inequality 4.1.35
in~\cite{AbrSte72}:
\[\forall x:~0< x \le 0.5828 \impLIES |\ln(1-x)| < \frac{3x}{2}.\]
The key to prove this inequality is to prove that the function 
\begin{eqnarray*}
G(x)&=&\frac{3x}{2} - \ln(1-x)
\end{eqnarray*}
satisfies $G(0.5828) > 0$. In PVS:
\begin{alltt}\small
  G(x|x < 1): real = 3*x/2 - ln(1-x)
   
  A_and_S : lemma G(0.5828) > 0

%|- A_and_S : PROOF (numerical :defs "G") QED
 \end{alltt}
In this case, the optional parameter \texttt{:defs "G"} tells
\texttt{numerical} that the user-defined function \texttt{G} has to be
expanded before performing the numerical evaluation. The original
proof of this lemma in PVS required the manual expansion of 19 terms
of the $\ln$ series.

The \texttt{numerical} strategy
is aimed to practicality rather than completeness.
In particular, it always terminate and it is configurable for
better accuracy (at the expense of performance).

Termination is trivially achieved as the strategy does not iterate for
different approximations, {\em i.e.}, step~\ref{go.eval} either goes to
step~\ref{go.next} or fails. In other words, if \texttt{numerical}
does not succeed, it does nothing. Furthermore, \texttt{numerical}
uses a default approximation parameter $n = 3$, which gives an
accuracy of about $2$ decimals for trigonometric functions. However,
the user can increase this parameter or set a different approximation
to each function according to his/her accuracy needs and availability of
computational power. Currently, there is no direct relation between
the approximation parameter and the accuracy, as all the
bounding functions have different convergence rates. On-going work
aims to provide, an absolute error of at most $2^{-p}$  for any expression with a new
approximation parameter $p$.
The strategy has not been designed to reuse
past computations.  Therefore, it will be prohibitively expensive to
automatically iterate \texttt{numerical} to achieve a small
approximation on a complex arithmetic expression.

In order to reduce the dependency effect, the \texttt{numerical} strategy 
automatically rearranges arithmetic
expressions using a simple factorization algorithm.  Due to the
subdistributivity property, the evaluation of factorized interval expressions 
is more accurate than that of non-factorized ones.
A set of lemmas of the NASA Langley PVS Libraries are also used as
rewriting rules on arithmetic expressions prior to numerical
evaluations.  This set of lemmas is parameterizable and can be
extended by the user.  For instance, trigonometric functions applied
to notable angles are automatically rewritten to their exact value.
Therefore, \texttt{numerical} is able to prove that $\sin(\pi/2) \in \I{1}$,
even if this proposition is not provable using our interval arithmetic operators alone.
Although it is not currently implemented, this approach can also be used
to normalize angles to the range $[-\pi,\pi]$ that is suitable
for the interval trigonometric functions in Sections~\ref{sec.trig}.

The splitting technique is implemented by allowing the user to specify
the number of tiles to be considered for each interval variable or
a default value for all of them. The strategy will evenly divide each interval.
For example, the simple expression in Section~\ref{sec.taylor} can be
proven to be in the range $\C{0,9/32}$ using a splitting of 16 subintervals.
\begin{alltt}\small
  fair : LEMMA
    x ## [|0,1|] IMPLIES x*(1-x) ## [|0,9/32|]

%|- fair : PROOF (instint :splitting 16) QED
\end{alltt}
In this example we have used the \texttt{instint} strategy. This strategy
is built on top of \texttt{numerical} and performs some
basic logic manipulations such as introduction of real variables 
and interval constants. In this case, 
the proof command \texttt{(initint :splitting 16)}
is equivalent to \texttt{(then (skeep) (numerical :vars ("x" "[|0,1|]" 16)))}.
It instructs PVS to introduce the real variable \texttt{x} and then 
to apply \texttt{numerical} by splitting 16 times the interval  $\C{0,1}$.

The taylor series expansion technique is implemented in two steps. First,
the \texttt{taylor} strategy automatically proves 
Proposition~\ref{prop.taylor} for a particular function $f$ and degree $n$. 
In the following example, we show that 
$x \in \I{x} \vdash x \times (1 - x) \in \sum_{i=0}^{2}(\I{x}_i \times (\I{x}-c)^i)/i!$, provided that $\I{x}$ is strictly proper.
\begin{alltt}\small
  F(X)  : MACRO Interval = X*(1-X)
  DF(X) : MACRO Interval = 1 - 2*X
  D2F(X): MACRO Interval = [| -2 |]

  ftaylor : LEMMA
    x ## X AND StrictlyProper?(X) IMPLIES
    x*(1-x) ## Taylor2[X](F,DF,D2F)

%|- ftaylor : PROOF (taylor) QED
\end{alltt}
The keyword \texttt{MACRO} tells the theorem prover 
to automatically expand the definition of the function. The expression
\texttt{Taylor2[X](F,DF,D2F)} corresponds to 
$\sum_{i=0}^{2}(\I{x}_i \times (\I{x}-c)^i)/i!$, where
\texttt{F}, \texttt{DF}, and \texttt{D2F} are the interval functions 
corresponding to $f$, it 1st, and its 2nd derivative.

Finally, the strategy \texttt{instint} 
is called with the lemma \texttt{ftaylor}.
\begin{alltt}\small
  best : LEMMA
    x ## [|0,1|] IMPLIES x*(1-x) ## [|0, 1/4|]

%|- best : PROOF 
%|-   (instint :taylor "ftaylor") 
%|- QED
\end{alltt}

\subsection{A simple case study}
\label{sec.div}
The arctangent function is heavily used in aeronautic applications as it is
fundamental to many Geodesic formulas\footnote{See, for example, 
Ed William's Aviation
Formulary at \url{http://williams.best.vwh.net/avform.htm}.}.
One common implementation technique uses an
approximation of the arctangent on the interval $\I{x} = \C{-1/30, 1/30}$
after argument reduction~\cite{Mar2K}.
For efficiency reasons, one may want to approximate the function
$\atan(x)$ to single precision by the polynomial
\begin{eqnarray*}
r(x) &=&  x - \frac{11184811}{33554432}x^3 - \frac{13421773}{67108864} x^5.
\end{eqnarray*}
The coefficients of the polynomial approximation are stored exactly
using IEEE single precision. 

The objective of this case study is to show 
that 
\[
x \in \C{-1/30,1/30}\ \vdash\ \atan(x) - r(x) \in \C{-2^{-i},2^{-i}}, 
\]
for different values of $i$. The PVS specification of this problem for
some values of $i$ is presented in Listing~\ref{lst.fairatan}. All the
lemmas are automatically discharged by the \texttt{instint} strategy
with different splitting and taylor's expansion degrees. As expected
taylor's expansions and splitting get better results than splitting
alone. Moreover, second degree expansions are almost always better
than first degree expansions. This is not necessarily the case as
illustrated by lemmas \texttt{fair\_atan\_t1\_14} and
\texttt{fair\_atan\_t2\_14}: for $i=14$, a first degree expansion with
no splitting is enough to prove the property, while a second degree
expansion requires a splitting of 2.

On a tile $\I{t}$ of $\I{x}$, the width of the error expression {\tt
  E} that does not use taylor's theorem evaluated on $\I{t}$ is larger
than the sum of the width of expressions {\tt Atan} and {\tt R}. As
the derivative of the arctangent is between $0.9989$ and $1$ on
$\I{x}$, we could expect that the width of {\tt R} is at least twice
the width of tile $\I{t}$. Therefore, to obtain an error
bound of $\C{-2^{-i},2^{-i}}$ we cannot use tiles larger than $2^{-i}$
and we will need at least $2^i/15 \approx 2^{i - 1.4}$ tiles.

We use the same kind of simple calculation to show that since $|e'(x)|
\le 2.37 \cdot 10^{-6}$ we will need about $2^{i - 14.8}$ tiles of
width $2^{-i} \cdot 10^{6} / 2.37$.  This figures are accurate when
we use second degree expansion but actual computations may require more
tiles due to some  dependency effects introduced when we use first
degree expansions.

\begin{algorithm*}
  \caption{Accuracy of the arctangent approximation}
  \label{lst.fairatan}
   \begin{alltt}\small

fair_atan : THEORY
BEGIN

    x    : var   real
    r(x) : MACRO real = x - (11184811/33554432) * x^3 - (13421773/67108864) * x^5
    e(x) : MACRO real = atan(x) - r(x)
    Xt   : Interval   = [| -1/30, 1/30 |]

    fair_atan_8 : LEMMA  x ## Xt IMPLIES e(x) ## [|-2^-8, 2^-8|]
%|- fair_atan_8 : PROOF (instint :splitting 18) QED

    X     : var   Interval
    R(X)  : MACRO Interval = X - 11184811/33554432 * X^3 - 13421773/67108864 * X^5
    E(X)  : MACRO Interval = Atan(X,4) - R(X)
    DE(X) : MACRO Interval = 
      1 / (1 + Sq(X)) - 1 + 3*(X^2*(11184811/33554432)) + 5*(X^4*(13421773/67108864))

    atan_taylor1 : LEMMA StrictlyProper?(X) AND x ## X IMPLIES e(x) ## Taylor1[X](E,DE)
%|- atan_taylor1 : PROOF (taylor) QED
    fair_atan_t1_14: LEMMA x ## Xt IMPLIES e(x) ## [|-2^-14, 2^-14|]
%|- fair_atan_t1_14 : PROOF (instint :taylor "atan_taylor1") QED
   fair_atan_t1_20: LEMMA x ## Xt IMPLIES e(x) ## [|-2^-20, 2^-20|]
%|- fair_atan_t1_20 : PROOF (instint :taylor "atan_taylor1" :splitting 13) QED

    D2E(X) : MACRO Interval =
      -2*X/Sq(1 + Sq(X)) + 20*(X^3*(13421773/67108864)) + 6*((11184811/33554432)*X)
     
    atan_taylor2 : LEMMA StrictlyProper?(X) AND x ## X IMPLIES e(x) ## Taylor2[X](E,DE,D2E)
%|- atan_taylor2 : PROOF (taylor) QED
    fair_atan_t2_14: LEMMA x ## Xt IMPLIES e(x) ## [|-2^-14, 2^-14|]
%|- fair_atan_t2_14 : PROOF (instint :taylor "atan_taylor2" :spitting 2) QED
    fair_atan_t2_20: LEMMA x ## Xt IMPLIES e(x) ## [|-2^-20, 2^-20|]
%|- fair_atan_t2_20 : PROOF (instint :taylor "atan_taylor2" :splitting 5) QED

END fair_atan
\end{alltt}
\end{algorithm*}

\begin{figure}
\[\includegraphics{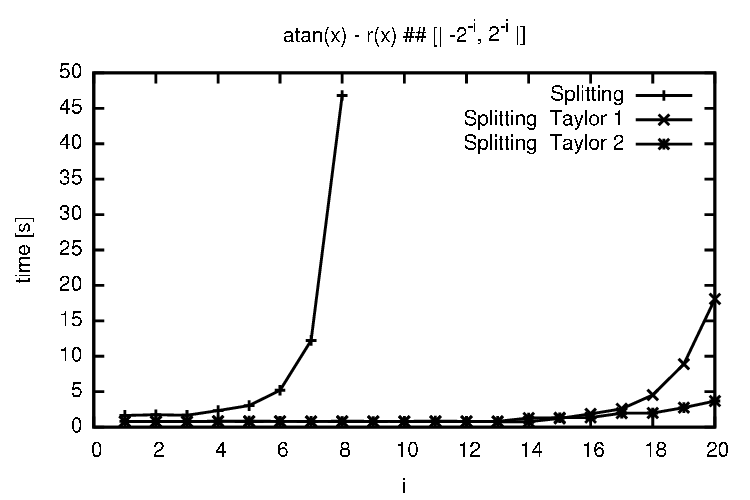}\]
\caption{Time required to prove $\tan(x)-r(x) \in \C{-1/30,1/30}$}
\label{fig.graph}
\end{figure}

Figure~\ref{fig.graph} presents a summary of the time required to
prove $\tan(x)-r(x) \in \C{-1/30,1/30}$ for $i$ in the range $[0,20]$
using splitting, splitting and first degree taylor's expansion, and
splitting and second degree taylor's expansion.

\subsection{Implementation and Performance Issues}

Actual definitions
in PVS have been slightly modified for efficiency reasons. 
For instance, multiplication is defined using a case analysis on the sign of 
the operands. Additionally, all interval operations are completed by returning
an empty interval if side conditions are not satisfied. This technique avoids 
some type correctness checks that are expensive.

The strategies in this library work over the 
PVS built-in real numbers. The major advantage of this approach is 
that the functionality of the strategies can be
extended to handle user defined real functions 
without modifying the strategy code. Indeed, optional parameters to
the \texttt{numerical} strategy allow for the specification of 
arbitrary real functions. If the interval interpretations are not provided,
the strategy tries to build them from the syntactic definition of the 
functions.
The trade-off for the use of the PVS type \texttt{real}, in
favor of a defined data type for arithmetic expressions, is that the
function $\Gn{e}$ and Theorem~{\ref{theo.incl}} are at the
meta-level, {\em i.e.}, they are not written in PVS. It also means that
the soundness of our method cannot be proven in PVS itself. In particular,
Theorem~{\ref{theo.incl}} has to be proven for each particular
instance of $e$ and $\Gn{e}$. This is not a major drawback as,
in addition to \texttt{numerical}, we
have developed a strategy called \texttt{inclusion} that 
discharges the sequent $\Gamma\ \vdash\ e \in \Gn{e}$ whenever is
needed. PVS strategies are conservative 
in the sense that they do not 
add inconsistencies to the theorem prover. Therefore, if \texttt{numerical}
succeeds to discharge a particular goal the answer is correct.

Finally, our method relies on explicit calculations to evaluate interval
expressions.  In theorem provers, explicit calculations usually means
symbolic evaluations, which are extremely inefficient for the interval
functions that we want to calculate.  
To avoid symbolic evaluations,
\texttt{numerical} is implemented using computational
reflection~\cite{har95,Bou97,vHPP98}. Interval expressions are
translated to Common Lisp (the implementation language of PVS) and
evaluated there. The extraction and evaluation mechanism is provided
by the PVS ground evaluator~\cite{Sha99}. The result of the evaluation
is translated back to the PVS theorem prover using the PVSio library
developed by one of the authors~\cite{Mun03NIA}.

\section{Conclusion and Limits of Tractability}

We have presented a pragmatic approach to verify ordinary real number 
computations in theorem provers. To this end, bounds for 
non-algebraic functions were established based on provable properties of
their approximation series. Furthermore, a library for interval arithmetic
was developed. The library includes strategies that automatically 
discharges numerical inequalities and interval inclusions.

The PVS Interval library contains 306 lemmas  in total.
It is roughly 10 thousand lines of specification and proofs and 1 thousand 
lines
of strategy definitions. These numbers do not take into account the bounding 
functions, which have been fully integrated to the NASA Langley PVS Libraries. 
It is difficult to estimate the human effort for this development as it has 
evolved over the years from an original axiomatic specification to a fully
foundational set of theories. As far as we know, this is the most complete 
formalization within a theorem prover of an interval arithmetic that includes
non-algebraic functions.

Research on interval analysis and exact arithmetic is rich and
abundant (see for example~\cite{Kea96,GowLes2K,Men05}). The goal of
interval analysis is to compute an upper bound of the round-off error
in a computation performed using floating-point numbers. In contrast,
in an exact arithmetic framework, an accuracy is specified at the
beginning of the computation and the computation is performed in such
way that the final result respects this accuracy.

Real numbers and exact arithmetic is also a subject of increasing
interest in the theorem proving community. Pioneers in this area were
Harrison and Gamboa who, independently, developed extensive
formalizations of real numbers for HOL~\cite{Har98} and
ACL2~\cite{Gam99}. In Coq, an axiomatic definition of reals is given
in~\cite{May01}, and constructive definitions of reals are provided
in~\cite{CiaDiG06} and~\cite{HugNiq06}.  As real numbers are built-in in
PVS, there is not much meta-theoretical work on real numbers. However,
a PVS library of real analysis was originally developed by
Dutertre~\cite{Dut96}~and~currently being maintained and extended as
part of the NASA Langley PVS Libraries. An alternative real analysis
library is proposed in~\cite{Got02}.

Closer to our approach are the tools presented in
\cite{DauMel04}~and~\cite{DauMelMun05}.  These tools generate bounds on the
round-off errors of numerical programs, and formal proofs that these
bounds are correct. The formal proofs are proof scripts that can be
checked off-line using a proof assistant.

Our approach is different from previous works in that we focus on
automation and pragmatism. In simple words, our
practical contribution is a correct pocket calculator for real
number computations in formal proofs.
Thanks to all the previous
developments in theorem proving and real numbers, lemmas like 
Lemma~\texttt{tr35} and Lemma~\texttt{A\_and\_S}
are provable in HOL, ACL2, Coq, or PVS.  The Interval library 
make these proofs routine in PVS.

\begin{figure}
\[\includegraphics{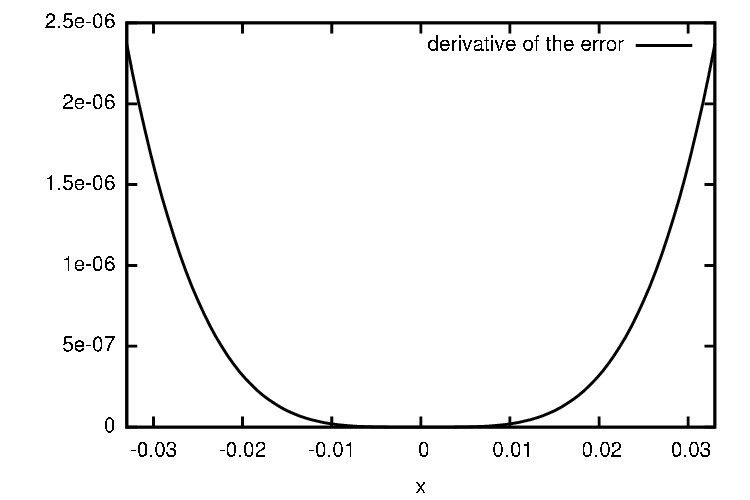}\]
\caption{Alternate {\tt fair\_atan} theorems will make use of interval arithmetic}
\label{fig.visible}
\end{figure}

As in real life, users benefit in managing both a pocket calculator
and a graphic tool. The fact that the example proposed in
Section~\ref{sec.div} is reaching the limits of tractability is not a
problem. Our library aims at providing some simple tools that can be
used seamlessly in proofs.  Figure~\ref{fig.visible} would prompt a
careful user that {\tt fair\_atan} theorems are a consequence of the
fact that the derivative of the error is always positive. Such a fact
could happen to be difficult to prove leading some one to prove that
the error is bounded on some subintervals and that the derivative is
always positive on some other subintervals. Anyways, such proofs will
involve our library more than once.

We continue developing this library and it is 
currently being used to check numerical properties of 
aircraft navigation algorithms developed at the National Institute of
Aerospace (NIA) and NASA. Future enhancements include:
\begin{itemize}
\item Development of a fully functional floating point arithmetic 
      library~\cite{BolMun06} in order to generate guaranteed proofs
      of round-off-errors~\cite{DauMel04}.
\item Integration of this library and an exact 
      arithmetic formalization in PVS developed by one of the 
      authors~\cite{LG02}.
\item Implementation of latest developments on Taylor
      Models~\cite{MakBer03,ChaDau06,ChaDauMunRev07}, which will enable 
      a greater automation of the Taylor's series expansion technique.
\end{itemize}

\bibliographystyle{IEEEtran}
\bibliography{daumas_ref,more}

\begin{thebibliography}{10}
\providecommand{\url}[1]{#1}
\csname url@rmstyle\endcsname
\providecommand{\newblock}{\relax}
\providecommand{\bibinfo}[2]{#2}
\providecommand\BIBentrySTDinterwordspacing{\spaceskip=0pt\relax}
\providecommand\BIBentryALTinterwordstretchfactor{4}
\providecommand\BIBentryALTinterwordspacing{\spaceskip=\fontdimen2\font plus
\BIBentryALTinterwordstretchfactor\fontdimen3\font minus
  \fontdimen4\font\relax}
\providecommand\BIBforeignlanguage[2]{{%
\expandafter\ifx\csname l@#1\endcsname\relax
\typeout{** WARNING: IEEEtran.bst: No hyphenation pattern has been}%
\typeout{** loaded for the language `#1'. Using the pattern for}%
\typeout{** the default language instead.}%
\else
\language=\csname l@#1\endcsname
\fi
#2}}

\bibitem{USA.92}
\BIBentryALTinterwordspacing
{Information Management and Technology Division}, ``Patriot missile defense:
  software problem led to system failure at {D}hahran, {S}audi {A}rabia,''
  United States General Accounting Office, Report B-247094, 1992. [Online].
  Available: \url{http://www.fas.org/spp/starwars/gao/im92026.htm}
\BIBentrySTDinterwordspacing

\bibitem{Lio.96}
\BIBentryALTinterwordspacing
J.-L. Lions \emph{et~al.}, ``Ariane 5 flight 501 failure report by the inquiry
  board,'' European Space Agency, Paris, France, Tech. Rep., 1996. [Online].
  Available: \url{http://ravel.esrin.esa.it/docs/esa-x-1819eng.pdf}
\BIBentrySTDinterwordspacing

\bibitem{GagMcC04}
\BIBentryALTinterwordspacing
D.~Gage and J.~McCormick, ``We did nothing wrong,'' \emph{Baseline}, vol.~1,
  no.~28, pp. 32--58, 2004. [Online]. Available:
  \url{http://common.ziffdavisinternet.com/download/0/2529/Baseline0304-Dissec%
tionNEW.pdf}
\BIBentrySTDinterwordspacing

\bibitem{Har98}
J.~Harrison, \emph{Theorem Proving with the Real Numbers}.\hskip 1em plus 0.5em
  minus 0.4em\relax Springer-Verlag, 1998.

\bibitem{FlePau2K}
\BIBentryALTinterwordspacing
J.~Fleuriot and L.~Paulson, ``Mechanizing nonstandard real analysis,''
  \emph{LMS Journal of Computation and Mathematics}, vol.~3, pp. 140--190,
  2000. [Online]. Available: \url{http://www.lms.ac.uk/jcm/3/lms1999-027/}
\BIBentrySTDinterwordspacing

\bibitem{Gam99}
\BIBentryALTinterwordspacing
R.~Gamboa, ``Mechanically verifying real-valued algorithms in {ACL2},'' Ph.D.
  dissertation, University of Texas at Austin, 1999. [Online]. Available:
  \url{ftp://ftp.cs.utexas.edu/pub/boyer/diss/gamboa.pdf}
\BIBentrySTDinterwordspacing

\bibitem{May01}
\BIBentryALTinterwordspacing
M.~Mayero, ``Formalisation et automatisation de preuves en analyse r{\'e}elle
  et num{\'e}rique,'' Ph.D. dissertation, Universit{\'e} Pierre et Marie Curie,
  Paris, France, 2001. [Online]. Available:
  \url{http://www.pps.jussieu.fr/~mayero/specif/these-mayero.ps}
\BIBentrySTDinterwordspacing

\bibitem{Got02}
\BIBentryALTinterwordspacing
H.~Gottliebsen, ``Automated theorem proving for mathematics: real analysis in
  {PVS},'' Ph.D. dissertation, University of St Andrews, 2001. [Online].
  Available: \url{http://www.dcs.qmul.ac.uk/~hago/thesis.ps.gz}
\BIBentrySTDinterwordspacing

\bibitem{MunCarDowBut03}
\BIBentryALTinterwordspacing
C.~Mu{\~n}oz, V.~Carre{\~n}o, G.~Dowek, and R.~Butler, ``Formal verification of
  conflict detection algorithms,'' \emph{International Journal on Software
  Tools for Technology Transfer}, vol.~4, no.~3, pp. 371--380, 2003. [Online].
  Available: \url{http://dx.doi.org/10.1007/s10009-002-0084-3}
\BIBentrySTDinterwordspacing

\bibitem{DauMelMun05}
\BIBentryALTinterwordspacing
M.~Daumas, G.~Melquiond, and C.~Mu{\~n}oz, ``Guaranteed proofs using interval
  arithmetic,'' in \emph{Proceedings of the 17th Symposium on Computer
  Arithmetic}, P.~Montuschi and E.~Schwarz, Eds., Cape Cod, Massachusetts,
  2005, pp. 188--195. [Online]. Available:
  \url{http://hal.archives-ouvertes.fr/hal-00164621}
\BIBentrySTDinterwordspacing

\bibitem{MunLes05}
\BIBentryALTinterwordspacing
C.~Mu{\~n}oz and D.~Lester, ``Real number calculations and theorem proving,''
  in \emph{18th International Conference on Theorem Proving in Higher Order
  Logics}, Oxford, England, 2005, pp. 239--254. [Online]. Available:
  \url{http://dx.doi.org/10.1007/11541868_13}
\BIBentrySTDinterwordspacing

\bibitem{OwrRusSha92}
\BIBentryALTinterwordspacing
S.~Owre, J.~M. Rushby, and N.~Shankar, ``{PVS}: a prototype verification
  system,'' in \emph{11th International Conference on Automated Deduction},
  D.~Kapur, Ed.\hskip 1em plus 0.5em minus 0.4em\relax Saratoga, New-York:
  Springer-Verlag, 1992, pp. 748--752. [Online]. Available:
  \url{http://pvs.csl.sri.com/papers/cade92-pvs/cade92-pvs.ps}
\BIBentrySTDinterwordspacing

\bibitem{AbrSte72}
M.~Abramowitz and I.~A. Stegun, Eds., \emph{Handbook of mathematical functions
  with formulas, graphs, and mathematical tables}.\hskip 1em plus 0.5em minus
  0.4em\relax Dover publications, 1972, ninth printing.

\bibitem{Mul06}
\BIBentryALTinterwordspacing
J.-M. Muller, \emph{Elementary functions, algorithms and implementation}.\hskip
  1em plus 0.5em minus 0.4em\relax Birkhauser, 2006. [Online]. Available:
  \url{http://www.springer.com/west/home/birkhauser/computer+science?SGWID=4-4%
0353-22-72377986-0}
\BIBentrySTDinterwordspacing

\bibitem{Neu90}
A.~Neumaier, \emph{Interval methods for systems of equations}.\hskip 1em plus
  0.5em minus 0.4em\relax Cambridge University Press, 1990.

\bibitem{JauKieDidWal01}
\BIBentryALTinterwordspacing
L.~Jaulin, M.~Kieffer, O.~Didrit, and E.~Walter, \emph{Applied interval
  analysis}.\hskip 1em plus 0.5em minus 0.4em\relax Springer, 2001. [Online].
  Available:
  \url{http://www.springeronline.com/sgw/cda/frontpage/0,10735,5-40106-22-2093%
571-0,00.html}
\BIBentrySTDinterwordspacing

\bibitem{Kea96}
R.~B. Kearfott, Ed., \emph{Rigorous global search: continuous problemes}.\hskip
  1em plus 0.5em minus 0.4em\relax Kluwer Academic Publishers, 1996.

\bibitem{Yak92}
A.~Yakovlev, ``Classification approach to programming of localizational
  (interval) computations,'' \emph{Interval Computations}, vol.~1, no.~1, pp.
  61--84, 1992.

\bibitem{Saw02}
J.~Sawada, ``Formal verification of divide and square root algorithms using
  series calculation,'' in \emph{3rd International Workshop on the ACL2 Theorem
  Prover and its Applications}.\hskip 1em plus 0.5em minus 0.4em\relax
  University of Grenoble, 2002, pp. 31--49.

\bibitem{Mun07NIA}
C.~Mu{\~{n}}oz, ``Batch proving and proof scripting in {PVS},'' NIA-NASA
  Langley, National Institute of Aerospace, Hampton, VA, Report NIA Report No.
  2007-03, NASA/CR-2007-214546, February 2007.

\bibitem{Mar2K}
P.~Markstein, \emph{{IA}-64 and elementary functions: speed and
  precision}.\hskip 1em plus 0.5em minus 0.4em\relax Prentice Hall, 2000.

\bibitem{har95}
J.~Harrison, ``Metatheory and reflection in theorem proving: A survey and
  critique,'' SRI Cambridge, Millers Yard, Cambridge, UK, Technical Report
  CRC-053, 1995.

\bibitem{Bou97}
S.~Boutin, ``Using reflection to build efficient and certified decision
  procedures,'' in \emph{Proceedings of the Third International Symposium on
  Theoretical Aspects of Computer Software}, London, United Kingdom, 1997, pp.
  515--529.

\bibitem{vHPP98}
F.~W. von Henke, S.~Pfab, H.~Pfeifer, and H.~Rue{\ss}, ``{Case Studies in
  Meta-Level Theorem Proving},'' in \emph{Proc. Intl. Conf. on Theorem Proving
  in Higher Order Logics}, ser. Lecture Notes in Computer Science, J.~Grundy
  and M.~Newey, Eds., no. 1479.\hskip 1em plus 0.5em minus 0.4em\relax
  Springer-Verlag, Sept. 1998, pp. 461--478.

\bibitem{Sha99}
N.~Shankar, ``Efficiently executing {PVS},'' Computer Science Laboratory, SRI
  International, Menlo Park, CA, Project report, Nov. 1999, available at
  \url{http://www.csl.sri.com/shankar/PVSeval.ps.gz}.

\bibitem{Mun03NIA}
C.~Mu{\~{n}}oz, ``Rapid prototyping in {PVS},'' NIA-NASA Langley, National
  Institute of Aerospace, Hampton, VA, Report NIA Report No. 2003-03,
  NASA/CR-2003-212418, May 2003.

\bibitem{GowLes2K}
\BIBentryALTinterwordspacing
P.~Gowland and D.~Lester, ``A survey of exact arithmetic implementations,'' in
  \emph{4th International Workshop on Computability and Complexity in
  Analysis}, Swansea, United Kingdom, 2000, pp. 30--47. [Online]. Available:
  \url{http://www.link.springer.de/link/service/series/0558/bibs/2064/20640030%
.htm}
\BIBentrySTDinterwordspacing

\bibitem{Men05}
\BIBentryALTinterwordspacing
V.~Ménissier-Morain, ``Arbitrary precision real arithmetic: design and
  algorithms,'' \emph{Journal of Logic and Algebraic Programming}, vol.~64,
  no.~1, pp. 13--39, 2005. [Online]. Available:
  \url{http://dx.doi.org/10.1016/j.jlap.2004.07.003}
\BIBentrySTDinterwordspacing

\bibitem{CiaDiG06}
\BIBentryALTinterwordspacing
A.~Ciaffaglione and P.~Di~Gianantonio, ``A certified, corecursive
  implementation of exact real numbers,'' \emph{Theoretical Computer Science},
  vol. 351, no.~1, pp. 39--51, 2006. [Online]. Available:
  \url{http://dx.doi.org/10.1016/j.tcs.2005.09.061}
\BIBentrySTDinterwordspacing

\bibitem{HugNiq06}
\BIBentryALTinterwordspacing
J.~Hughes and M.~Niqui, ``Admissible digit sets,'' \emph{Theoretical Computer
  Science}, vol. 351, no.~1, pp. 61--73, 2006. [Online]. Available:
  \url{http://dx.doi.org/10.1016/j.tcs.2005.09.059}
\BIBentrySTDinterwordspacing

\bibitem{Dut96}
\BIBentryALTinterwordspacing
B.~Dutertre, ``Elements of mathematical analysis in {PVS},'' in \emph{Theorem
  Proving in Higher Order Logics: 9th International Conference. TPHOLs'97},
  ser. Lecture Notes in Computer Science, J.~von Wright, J.~Grundy, , and
  J.~Harrison, Eds., no. 1125.\hskip 1em plus 0.5em minus 0.4em\relax Turku,
  Finland: Springer-Verlag, August 1996, pp. 141--156. [Online]. Available:
  \url{http://www.sdl.sri.com/papers/tphol96/}
\BIBentrySTDinterwordspacing

\bibitem{DauMel04}
\BIBentryALTinterwordspacing
M.~Daumas and G.~Melquiond, ``Generating formally certified bounds on values
  and round-off errors,'' in \emph{Real Numbers and Computers}, Dagstuhl,
  Germany, 2004, pp. 55--70. [Online]. Available:
  \url{http://hal.inria.fr/inria-00070739}
\BIBentrySTDinterwordspacing

\bibitem{BolMun06}
\BIBentryALTinterwordspacing
S.~Boldo and C.~Muñoz, ``Provably faithful evaluation of polynomials,'' in
  \emph{Proceedings of the 2006 ACM Symposium on Applied Computing}, Dijon,
  France, 2006, pp. 1328--1332. [Online]. Available:
  \url{http://doi.acm.org/10.1145/1141277.1141586}
\BIBentrySTDinterwordspacing

\bibitem{LG02}
D.~Lester and P.~Gowland, ``Using {PVS} to validate the algorithms of an exact
  arithmetic,'' \emph{Theoretical Computer Science}, vol. 291, no.~2, pp.
  203--218, Nov. 2002.

\bibitem{MakBer03}
\BIBentryALTinterwordspacing
K.~Makino and M.~Berz, ``Taylor models and other validated functional inclusion
  methods,'' \emph{International Journal of Pure and Applied Mathematics},
  vol.~4, no.~4, pp. 379--456, 2003. [Online]. Available:
  \url{http://bt.pa.msu.edu/pub/papers/TMIJPAM03/TMIJPAM03.pdf}
\BIBentrySTDinterwordspacing

\bibitem{ChaDau06}
\BIBentryALTinterwordspacing
F.~Ch\'aves and M.~Daumas, ``A library to {T}aylor models for {PVS} automatic
  proof checker,'' in \emph{Proceedings of the NSF workshop on reliable
  engineering computing}, Savannah, Georgia, 2006, pp. 39--52. [Online].
  Available:
  \url{http://www.gtsav.gatech.edu/workshop/rec06/papers/Chaves_paper.pdf}
\BIBentrySTDinterwordspacing

\bibitem{ChaDauMunRev07}
\BIBentryALTinterwordspacing
F.~Cháves, M.~Daumas, C.~Muñoz, and N.~Revol, ``Automatic strategies to
  evaluate formulas on {T}aylor models and generate proofs in {PVS},'' in
  \emph{6th International Congress on Industrial and Applied Mathematics},
  Zurich, Switzerland, 2007. [Online]. Available: \url{http://www.iciam07.ch/}
\BIBentrySTDinterwordspacing

\end{thebibliography}

\end{document}